# A practical illustration of the importance of realistic individualized treatment rules in causal inference

**Oliver Bembom and Mark J. van der Laan**

*University of California at Berkeley*
*Division of Biostatistics*
*School of Public Health*
*101 Haviland Hall #7358*
*Berkeley, California 94720-7358*
*e-mail:* `bembom@berkeley.edu`; `laan@stat.berkeley.edu`

**Abstract:** The effect of vigorous physical activity on mortality in the elderly is difficult to estimate using conventional approaches to causal inference that define this effect by comparing the mortality risks corresponding to hypothetical scenarios in which all subjects in the target population engage in a given level of vigorous physical activity. A causal effect defined on the basis of such a static treatment intervention can only be identified from observed data if all subjects in the target population have a positive probability of selecting each of the candidate treatment options, an assumption that is highly unrealistic in this case since subjects with serious health problems will not be able to engage in higher levels of vigorous physical activity. This problem can be addressed by focusing instead on causal effects that are defined on the basis of realistic individualized treatment rules and intention-to-treat rules that explicitly take into account the set of treatment options that are available to each subject. We present a data analysis to illustrate that estimators of static causal effects in fact tend to overestimate the beneficial impact of high levels of vigorous physical activity while corresponding estimators based on realistic individualized treatment rules and intention-to-treat rules can yield unbiased estimates. We emphasize that the problems encountered in estimating static causal effects are not restricted to the IPTW estimator, but are also observed with the *G*-computation estimator, the DR-IPTW estimator, and the targeted MLE. Our analyses based on realistic individualized treatment rules and intention-to-treat rules suggest that high levels of vigorous physical activity may confer reductions in mortality risk on the order of 15-30%, although in most cases the evidence for such an effect does not quite reach the 0.05 level of significance.

## Contents

## 1. Introduction

A substantial body of epidemiologic research indicates that recent and current physical activity in the elderly are associated with reductions in cardiovascular morbidity and mortality and improvement in or prevention of metabolic abnormalities that place elderly people at risk for these outcomes [2; 5; 8; 21; 27]. Based on these findings, the CDC currently recommends that elderly people engage in moderate-intensity physical activities such as bicycling on level terrain for 30 minutes or more at least five times a week in order to maintain their health [3].

While epidemiologic studies have produced compelling evidence for the health benefits provided by such moderate-intensity physical activities, it remains a largely open question to what extent more vigorous physical activities can offer additional benefits to the elderly. One of the main reasons for why this question has proven difficult to investigate lies in the difficulties encountered by conventional statistical methods for causal inference in this context. These methods would typically define the causal effect of vigorous physical activity on a health outcome of interest by comparing the distribution of that outcome under the hypothetical scenario in which all subjects in the target population exercise at a given activity level to the corresponding distribution under the reference scenario in which all subjects abstain from vigorous physical activity. In order to estimate such treatment-specific counterfactual outcome distributions from observational data, however, one has to assume not only that the investigator has recorded all relevant confounding factors, but also that all subjects in the target population have a positive probability of selecting each of the treatment levels under consideration. Intuitively, this latter positivity assumption, also referred to as the assumption of experimental treatment assignment (ETA), makes sense since we should not be able to estimate the counterfactual outcome distribution corresponding to a given treatment level if there exists a subgroup of the target population that in reality is never observed at that treatment level.

In the context of studying the benefits of vigorous physical activity in the elderly, this assumption appears highly unrealistic for two reasons. First, serious health problems would prevent a considerable proportion of subjects from ever participating in the highest level of vigorous physical activity. Since we could not devise an intervention under which such subjects would exercise at the highest level, the corresponding counterfactual outcomes, as first pointed out by Robins [13; 14], are not even well defined, making it in fact meaningless to talk about an outcome distribution we would observe if all subjects were assigned to the highest activity level [18]. Second, while it may be more reasonable to assume that all subjects could at least hypothetically participate in intermediate activity levels, we might still expect that there is a fair number of subjects in our target population that due to poor health would in reality only be observed at the very lowest activity levels. In the absence of strong additional modeling assumptions, such a violation of the ETA assumption would likley cause a conventional causal analysis to overestimate the beneficial impact of higher levels of vigorous exercise since any estimate of the corresponding counterfactual distribution would be based on a group of subjects that is healthier than the population as a whole [13; 14].

These two problems can be addressed by defining the causal effect of interest on the basis of interventions that, in contrast to the static interventions described above, explicitly take into account the set of treatment options available to each subject [13; 14]. In the present case, we might consider hypothetical scenarios in which subjects are assigned to a particular vigorous activity level unless they rate their own health as poor, in which case they will be assigned to the lowest activity level. The causal effect of vigorous physical activity could then be defined by comparing the outcome distribution we would observe for different target levels to the corresponding distribution we would observe under no vigorous physical activity. Both Inverse-Probability-of-Treatment-Weighted (IPTW) [15; 19] and double robust [20; 24] estimators for mean counterfactual outcomes corresponding to such individualized treatment rules have been proposed.

Recently such estimators have also been been proposed for two kinds of realistic interventions that, unlike in the example above, are not specified *a priori* by the investigator, but are instead defined implicitly on the basis of the observed data [12; 28]. These interventions make use of a subject's estimated conditional probability of selecting a particular treatment option, given baseline characteristic, to decide if that treatment option is realistic. Specifically, treatment options are considered unrealistic if this estimated probability falls below a user-supplied minimum level such as 0.05. The first intervention is based on realistic individualized treatment rules that assign a treatment level that is as close as possible to a specified target level while still being a realistic option for that subject. In the context of physical activity, for instance, we might consider rules that assign subjects to the highest vigorous activity level not exceeding a specified target level that they are still realistically capable of. The second intervention is based on intention-to-treat rules that, like realistic individualized treatment rules, attempt to assign subjects to a specified target level, but allow subjects for whom this target level is not realistic to follow their self-selected treatment level rather than assigning them to the next highest realistic level. Causal effect estimates based on such rules thus aim to produce the results of an intention-to-treat analysis of a randomized trial in which a proportion of subjects fail to comply with treatment assignment and instead select their own treatment level.

In this article, we present a data analysis examining the potential benefits of vigorous-intensity physical activity that compares the results obtained through a conventional analysis to those based on realistic indivualized treatment and intention-to-treat rules. Our analysis illustrates that a conventional analysis based on static treatment rules yields severely biased results that dramatically overestimate the true effect of higher levels of vigorous physical activity. At the same time, we show that causal effects based on realistic individualized treatment rules and intention-to-treat rules can be estimated without bias. The remainder of the article is organized as follows. After describing our data source, we briefly review the counterfactual framework for causal inference and describe the various estimators that have been proposed for estimating causal effects. We then present the details of our data analysis and close with a brief discussion of our results.

## 2. Data source

Tager et al. [26] followed a group of people aged 55 years and older living in and around Sonoma, CA, over a time period of about ten years as part of a community-based longitudinal study of physical activity and fitness (Study of Physical Performance and Age Related Changes in Sonomans - SPPARCS). Our goal in analyzing the data that were collected as part of this study is to examine the effect of vigorous LTPA as recorded at the baseline interview on subsequent five-year all-cause mortality.

Our measure of vigorous LTPA is defined based on a questionnaire in which participants were asked how many hours during the past seven days they had participated in twelve common vigorous physical activities such as jogging, swimming, bicycling on hills, or racquetball. Activities were assigned standard intensity values in metabolic equivalents (METs) [1]; one MET approximately equals the oxygen consumption required for sitting quietly. A continuous summary score was obtained by multiplying these intensity values by the number of hours engaged in the various activities and summing up over all activities considered here. The treatment variable $A$ was then defined as a categorical version of this summary LTPA score:

$$A = \begin{cases} 0 & \text{if } LTPA = 0 \text{ METs} \\ 1 & \text{if } 0 \text{ METs} < LTPA \leq 10 \text{ METs} \\ 2 & \text{if } 10 \text{ METs} < LTPA \leq 20 \text{ METs} \\ 3 & \text{if } 20 \text{ METs} < LTPA \leq 40 \text{ METs} \\ 4 & \text{if } 40 \text{ METs} < LTPA \leq 60 \text{ METs} \\ 5 & \text{if } 60 \text{ METs} < LTPA \end{cases} \quad (1)$$

To compare, the current CDC recommendation for engaging in moderate-intensity physical activity for 30 minutes at least five times a week corresponds to an energy expenditure of 22.5 METs.

Apart from sex and age, the primary confounding factor of the relationship between LTPA and all-cause mortality is likely to be given by a subject's underlying level of general health. Healthier subjects will not only tend to experience lower mortality risks, but are also more likely to engage in higher levels of vigorous physical activity. To control for this source of confounding, our analysis adjusts for a number of covariates that are intended to capture a subject's underlying level of health. Participants were asked, for instance, to rate their health as excellent, good, fair, or poor. Self-reported physical functioning was defined from a series of questions that assessed the degree of difficulty a participant experienced in various activities of daily living [9; 22]. On the basis of this questionnaire, we classified a participant's level of physical functioning as excellent, moderately impaired, or severely impaired. In addition, participants were asked about the previous occurrence of cardiac events such as myocardial infarctions, the presence of a number of chronic health conditions, their smoking status, as well as a possible decline in physical activity compared to 5 or 10 years earlier. Table 1 summarizes the definition of the covariates we adjust for as potential confounding factors.

TABLE 1
*Definition of indicator variables that are considered as potential confounders.*

| Variable | Definition |
| --- | --- |
| $FEMALE$ | Female |
| $AGE.1$ | $\leq 60$ years old |
| $AGE.2$ | 60-70 years old |
| $AGE.4$ | 80-90 years old |
| $AGE.5$ | 90-100 years old |
| $HTL.EX$ | Excellent self-rated health |
| $HLT.FAIR$ | Fair self-rated health |
| $HLT.POOR$ | Poor self-rated health |
| $NRB.FAIR$ | Moderately impaired physical functioning ($0.5 \leq$ NRB score ¡ 1.0) |
| $NRB.POOR$ | Severely impaired physical functioning (NRB score ¡ 0.5) |
| $CARD$ | Previous occurrence of any of the following cardiac events: Angina, myocardial infarction, congestive heart failure, coronary by-pass surgery, and coronary angioplasty |
| $CHRON$ | Presence of any of the following chronic health conditions: stroke, cancer, liver disease, kidney disease, Parkinson's disease, and diabetes mellitus |
| $SMK.CURR$ | Current smoker |
| $SMK.EX$ | Former smoker |
| $DECLINE$ | Activity decline compared to 5 or 10 years earlier |

Of the 2092 participants enrolled in the SPPARCS study, 15 did not answer all the questions needed to define their level of vigorous physical activity; an additional 26 were missing information about at a least one of the confounding factors described above. Our analysis is based on the remaining 2051 participants. We note that the outcome of interest, five-year survival status, was available for all study participants so that we do not have to adjust for right censoring.

## 3. Methods

The observed data are given by $n$ i.i.d. copies of $O = (W, A, Y)$, where $W$ denotes the collection of adjustment variables, $A$ gives the categorical physical activity level, and $Y$ is an indicator for death in the five years following the baseline interview. Within the counterfactual framework for causal inference, as first introduced by Neyman [11] and further developed by Rubin [23] and Robins [13; 14], this observed data structure $O$ is viewed as a censored version of a hypothetical full data structure $X = (Y_a : a \in \mathcal{A})$ that contains the outcome $Y_a$ we would have observed on this subject had she been assigned to treatment level $a$ for all $a$ in the collection $\mathcal{A} = \{0, 1, \ldots, 5\}$ of possible treatment levels. The causal effect of vigorous physical activity on all-cause mortality could now be defined by comparing the mortality risk $E[Y_a]$ we would observe if all subjects in the target population exercised at a given level $a > 0$ to the corresponding mortality risk $E[Y_0]$ we would observe if all subjects abstained from vigorous physical activity. As discussed above, this definition of a causal effect would require, however, that the counterfactual outcomes $Y_a$ are well defined for all subjects.

A mean counterfactual outcome $E[Y_a]$ can only be estimated from the observed data if the investigator has recorded all relevant confounding factors and if all subjects in the target population have positive probability of selecting each of the treatment levels. This latter assumption of experimental treatment assignment can be formalized by requiring that for all candidate static treatment interventions $a = 0, 1, \ldots, 5$, we have with probability 1.0 that

$$g(a \mid W) \equiv P(A = a \mid W) > 0. \tag{2}$$

In fact, it has been shown that estimation of mean counterfactual outcomes becomes problematic even if there exist values of $a$ and $W$ for which the treatment assignment probabilities $g(a \mid W)$ are not identically equal to zero, but very close to zero [10]. To avoid problems due to such a practical violation of the ETA assumption, we may hence require in practice that, for $a = 0, 1, \ldots, 5$, we have $g(a \mid W) > \alpha$ with probability 1.0, with $\alpha = 0.05$, for instance.

Estimators of causal effects defined on the basis of the realistic individualized treatment rules do not rely on the ETA assumption. Given a target treatment level $a$ and a subject's baseline covariates $W$, such rules assign the highest treatment level not exceeding $a$ that the subject is still realistically capable of. Specifically, let

$$\mathcal{D}(W) = \{a \in \mathcal{A} : g(a \mid W) \geq \alpha\} \tag{3}$$

denote the set of treatment options that, given baseline covariates $W$, are realistic for a particular subject in the sense that she would select any one of those treatment options with a probability of at least $\alpha$. A realistic individualized treatment rule can then be defined as

$$d(a, W) = \max\{a^* \in \mathcal{D}(W) : a^* \leq a\}. \tag{4}$$

As with static treatment regimens, we use the notation $Y_{d(a,W)}$ to denote the outcome we would have observed on the subject had she followed the individualized rule $d(a, W)$, i.e. $Y_{d(a,W)} \equiv Y_{\tilde{a}}$ where $\tilde{a} = d(a, W)$. A realistic causal effect of vigorous physical activity on all-cause mortality can now be defined by comparing the mortality risk $E[Y_{d(a,W)}]$ we would observe if all subjects in the target population followed a given rule $d(a, W)$, $a > 0$, to the corresponding mortality risk $E[Y_{d(0,W)}] = E[Y_0]$ we would observe if all subjects abstained from vigorous physical activity. By the definition of $d(a, W)$, we have, for $a = 0, 1, \ldots, 5$, that $g(d(a, W) \mid W) > \alpha$ with probability 1.0, demonstrating that the equivalent of assumption (2) is trivially satisfied in estimating the corresponding causal effects.

Under an intention-to-treat rule $d(a, A, W)$, subjects are assigned to a specified target treatment level $a$ if that treatment level represents a realistic option for them, but are allowed to follow their self-selected treatment $A$ otherwise:

$$d(a, A, W) = I(a \in \mathcal{D}(W))a + I(a \notin \mathcal{D}(W))A. \qquad (5)$$

An intention-to-treat causal effect of vigorous physical activity on all-cause mortality can now be defined by comparing the counterfactual mortality risks $E[Y_{d(a,A,W)}]$, $a > 0$, and $E[Y_{d(0,A,W)}] = E[Y_0]$. Note that we have

$$E[Y_{d(a,A,W)}] = E\Big[Y_a I(a \in \mathcal{D}(W))\Big] + E\Big[Y I(a \notin \mathcal{D}(W))\Big]. \qquad (6)$$

The second quantity is trivially identified by the observed data, and $a \in \mathcal{D}(W)$ guarantees that $g(a \mid W) > \alpha$ with probability 1.0, ensuring identifiability of the second quantity, so that the equivalent of assumption (2) is guaranteed to hold in the estimation of intention-to-treat causal effects. We note that the true treatment mechanism $g$ and therefore also the set $\mathcal{D}(W)$ of realistic treatment options will generally be unknown. In practice, it will therefore usually be necessary to substitute a given estimate $g^*$ of the treatment mechanism $g$ in the definition of $D(W)$.

Several different classes of estimators have been proposed for estimating mean counterfactual outcomes corresponding to static treatment rules: $G$-computation estimators [13], Inverse-Probability-of-Treatment-Weighted (IPTW) estimators [16], double robust IPTW (DR-IPTW) estimators [29], regression-like DR estimators [16] that were later recognized to be an example of the general class of targeted maximum-likelihood estimators [30], and the $d - d^*$ structural-nested-mean-model (SNMM) estimators [17]; all of these estimators have natural analogues in the context of realistic individualized treatment rules and intention-to-treat rules. While it is well known that the IPTW estimator can suffer from considerable bias if the ETA assumption is violated, the remaining four estimators are in fact also severely compromised in such situations in that they now have to rely fully on model assumptions that cannot be tested from the data [10]. Since this latter phenomenon is rarely discussed in the literature, we will provide a practical illustration by comparing the estimates obtained by the first four of these estimators for the three different causal effects defined above. We next review the definition and implementation of the four estimators of interest in order to be able to discuss their behavior in more detail.

We begin with estimators of the mean counterfactual outcome $\psi = E[Y_{d(a,W)}]$ for a given realistic individualized treatment rule $d(a, W)$. Note that the mean counterfactual outcome $E[Y_a]$ for a given static treatment rule corresponds to the special case of setting $\alpha = 0$ in the definition of $\mathcal{D}(W)$. The G-computation estimator of $\psi$ is based on the observation that under the assumption of no unmeasured confounders, this parameter is identified by the observed data as

$$\psi = E[Y_{d(a,W)}] = E_W \Big[ E[Y \mid A = d(a,W), W] \Big]. \tag{7}$$

This immediately implies a substitution estimator based on estimates of the marginal distribution of $W$, $P(W)$, and the conditional distribution of $Y$ given $A$ and $W$, $P(Y \mid A, W)$. The first distribution can be estimated non-parametrically by the empirical distribution of $W$ in our sample, but estimation of $P(Y \mid A, W)$ will generally require specification of a parametric model. In the case of a binary outcome $Y$, an estimate $Q_n$ of the regression $Q(A, W) = E[Y \mid A, W]$ based on an appropriate logistic regression model completely defines an estimate of the conditional distribution $P(Y \mid A, W)$. The corresponding substitution estimator for $\psi$ is then given by

$$\psi_n^{G-comp} = \frac{1}{n} \sum_{i=1}^{n} Q_n(d(a, W_i), W_i). \tag{8}$$

This estimator gives a consistent estimate of $\psi$ if the model for $Q(A, W)$ is correctly specified.

The IPTW and DR-IPTW estimators are based on a general estimating function methodology that is based on the following three steps [29]. First, estimating functions for $\psi$ are obtained assuming that we have access to the full data structure $X$. These estimating functions are then mapped into functions of the observed data structure by applying an IPTW mapping. Lastly, a class of more robust and efficient estimating functions is obtained by subtracting from these IPTW estimating functions their projection onto the tangent space for the treatment mechanism in the model that only makes the assumption of no unmeasured confounders. In a non-parametric model, the only unbiased full-data estimating function for $\psi$ is given by

$$D^{Full}(X \mid \psi) = Y_{d(a,W)} - \psi. \tag{9}$$

A corresponding IPTW estimating function is given by

$$D^{IPTW}(O \mid g, \psi) = \frac{I(A = d(a, W))}{g(A \mid W)} Y - \psi. \tag{10}$$

The IPTW estimator $\psi_n^{IPTW}$ is defined as the solution of the estimating equation

$$0 = \frac{1}{n} \sum_{i=1}^{n} D^{IPTW}(O_i \mid g_n, \psi), \tag{11}$$

where $g_n$ is an estimate of $g$ that may, for example, be obtained as the maximum-likelihood estimate of $g$ in an appropriately specified parametric model. Specifically, this estimator is given by

$$\psi_n^{IPTW} = \frac{1}{n} \sum_{i=1}^{n} \frac{I(A_i = d(a, W_i))}{g_n(A_i \mid W_i)} Y_i. \tag{12}$$

It gives a consistent estimate of $\psi$ if the model for the treatment mechanism $g$ is correctly specified.

The projection of $D^{IPTW}$ onto the nuisance tangent space $T_{NUC}$ corresponding to the treatment mechanism under the assumption of no unmeasured confounders can be computed as

$$\begin{aligned}\Pi[D^{IPTW} \mid T_{NUC}] &= E[D^{IPTW} \mid A, W] - E[D^{IPTW} \mid W]\\ &= \frac{I(A = d(a, W))}{g(A \mid W)} Q(A, W) - Q(d(a, W), W)\end{aligned}$$

so that the DR-IPTW estimating function is given by

$$D^{DR}(O \mid g, Q, \psi) = \frac{I(A = d(a, W))}{g(A \mid W)} \Big[Y - Q(A, W)\Big] + Q(d(a, W), W) - \psi. \tag{13}$$

The corresponding DR-IPTW estimator $\psi_n^{DR}$ is defined as the solution of the estimating equation

$$0 = \frac{1}{n} \sum_{i=1}^{n} D^{DR}(O_i \mid g_n, Q_n, \psi). \tag{14}$$

Specifically,

$$\psi_n^{DR} = \frac{1}{n} \sum_{i=1}^{n} \frac{I(A_i = d(a, W_i))}{g_n(A_i \mid W_i)} \Big[Y_i - Q_n(A_i, W_i)\Big] + Q_n(d(a, W_i), W_i). \tag{15}$$

This estimator gives a consistent estimate of $\psi$ if the model for either $g$ or $Q$ is correctly specified. It is also locally efficient in the sense that correct specification of both models yields an efficient estimator.

Like the $G$-computation estimator, the targeted MLE of $\psi$ is a substitution estimator based on estimates of the components $P(W)$ and $P(Y \mid A, W)$ of the observed data density. In order to avoid relying on an *a priori* specified parametric model for the latter component, we may often want to employ a data-adaptive model selection approach such as the Deletion/Substituion/Addition algorithm [25] or Least Angle Regression [4] for the purposes of estimating this conditional density. This is somewhat problematic, however, since such algorithms will select an appropriate model based on a criterion that is aimed at estimating the nuisance parameter $P(Y \mid A, W)$ efficiently, which in general does not lead to an efficient estimator of the parameter of interest $\psi$. The targeted MLE therefore first updates the initial estimate of the observed-data density that would be used by the *G*-computation estimator in a way that targets estimation of this density at the parameter of interest and makes the corresponding substitution estimator double robust and locally efficient. Specifically, this is achieved by formulating a parametric model indexed by a Euclidean parameter $\epsilon$ through the initial estimate of the observed-data density at $\epsilon = 0$ whose scores include the components of the efficient influence curve of $\psi$ at the initial density estimate, obtaining a maximum-likelihood estimate of $\epsilon$ in this model, and updating the original density estimate accordingly.

Since this targeted maximum-likelihood approach was only recently developed, we will illustrate it here in the context of estimating the parameter of interest $\psi$. For this purpose, let $P_n^0$ be an initial estimator of the observed-data density that estimates the marginal distribution of $W$ by the empirical distribution of $W$, the treatment mechanism $g$ by an estimate $g(P_n^0)$, and the conditional distribution of $Y$ given $A$ and $W$ by an initial fit $Q(P_n^0)$ that can be represented in the form of the logistic function

$$Q(P_n^0)(A, W) = \frac{1}{1 + \exp(-m_n^0(A, W))}. \tag{16}$$

We then need to formulate a parametric fluctuation through this initial density estimate whose scores at the initial estimate include the components of the efficient influence curve for $\psi$. This efficient influence curve, given by the influence curve $D(P)$ of the DR-IPTW estimator

$$D(P) = \frac{I(A = d(a, W))}{g(A \mid W)} \Big[ Y - Q(A, W) \Big] + Q(d(a, W), W) - \psi, \tag{17}$$

can be decomposed as

$$\begin{aligned} D(P) &= D(P) - E[D(P) \mid A, W] + \\ & \quad E[D(P) \mid A, W] - E[D(P) \mid W] + \\ & \quad E[D(P) \mid W] - E[D(P)], \end{aligned} \tag{18}$$

corresponding to scores for $P(Y \mid A, W)$, $P(A \mid W)$, and $P(W)$, respectively. Specifically, we have that

$$\begin{aligned} D_1(P) &= D(P) - E[D(P) \mid A, W] \\ &= \frac{I(A = d(a, W))}{g(A \mid W)} \Big[ Y - Q(A, W) \Big] \end{aligned} \tag{19}$$

$$\begin{aligned} D_2(P) &= E[D(P) \mid A, W] - E[D(P) \mid W] \\ &= 0 \end{aligned} \tag{20}$$

$$\begin{aligned} D_3(P) &= E[D(P) \mid W] - E[D(P)] \\ &= Q(d(a, W), W) - \psi_a. \end{aligned} \tag{21}$$

Since the empirical distribution of $W$ is a non-parametric maximum-likelihood estimator of $P(W)$, it in particular equals the MLE of $P(W)$ in any parametric fluctuation through this initial estimate so that we do not need to concern ourselves with updating this component of the observed-data density. Since the parameter of interest is orthogonal to the treatment mechanism $g$ so that $D_2(P) = 0$, we also do not need to obtain an update of an initial estimate of $g$. As a submodel through $P_n^0(Y \mid A, W)$, we will consider a logistic regression model that is identical to the initial fit $Q(P_n^0)$ except for an added covariate $h(P_n^0)(A, W)$:

$$Q(P_n^0)(\epsilon)(A, W) = \frac{1}{1 + \exp(-m_n^0(A, W) - \epsilon h(P_n^0)(A, W))} \quad (22)$$

The covariate $h(P_n^0)(A, W)$ needs to be chosen such that the score of this submodel at $\epsilon = 0$ is equal to $D_1(P_n^0)$, the component of the efficient influence curve corresponding to $P(Y \mid A, W)$ at the initial density estimate. The score of the selected submodel at $\epsilon = 0$ is given by

$$S(0) = h(P_n^0)(A, W)\Big(Y - Q(P_n^0)(A, W)\Big). \quad (23)$$

Solving for $h$ such that

$$\begin{aligned} S(0) &= D_1(P_n^0) \\ &= \frac{I(A = d(a, W))}{g(P_n^0)(A \mid W)}\Big[Y - Q(P_n^0)(A, W)\Big] \end{aligned} \quad (24)$$

yields the solution

$$h(P_n^0)(A, W) = \frac{I(A = d(a, W))}{g(P_n^0)(A \mid W)}. \quad (25)$$

Let $\epsilon_n$ denote the MLE of $\epsilon$ in $Q(P_n^0)(\epsilon)$, which can be obtained by simply regressing $Y$ on $h(P_n^0)(A, W)$ according to a logistic regression model with offset equal to $m_n^0(A, W)$. The targeted MLE of $\psi$ is then given by the substitution estimator based on the updated estimate

$$Q_n^1(A, W) = \frac{1}{1 + \exp(-m_n^0(A, W) - \epsilon_n h(P_n^0)(A, W))} \quad (26)$$

of the regression $Q(A, W)$. Specifically, we have that

$$\psi_n^{tMLE} = \frac{1}{n}\sum_{i=1}^{n} Q_n^1(d(a, W_i), W_i). \quad (27)$$

To summarize, implementing this estimator thus requires initial estimates of the regression $Q$ and the treatment mechanism $g$ as they would also be used by the three estimators described above, updating the estimate for $Q$ in a simple univariate logistic regression, and then computing the corresponding substitution estimator of $\psi$. The resulting targeted MLE solves the double robust estimating equation based on $Q_n^1(A, W)$ and $g_n$, i.e.

$$\frac{1}{n}\sum_{i=1}^{n} \frac{I(A_i = d(a, W_i))}{g(P_n^0)(A_i \mid W_i)}\Big[Y_i - Q_n^1(A_i, W_i)\Big] + Q_n^1(d(a, W_i), W_i) - \psi_n^{tMLE} = 0, \quad (28)$$

so that it is in fact equivalent to the DR-IPTW estimator given in (15) with $Q_n^1(A, W)$ substituted for $Q_n(A, W)$. Like the DR-IPTW estimator, the targeted MLE is therefore consistent if at least one of the two nuisance parameters $g$ and $Q$ is estimated consistently. Similarly, the estimator is locally efficient in the sense that it is efficient if both of these nuisance parameters are estimated consistently.

As mentioned previously, estimation of the mean counterfactual outcome $E[Y_a]$ corresponding to a static treatment intervention represents a special case of the realistic individualized treatment rules considered here. $G$-computation, IPTW, and DR-IPTW estimators of the mean counterfactual outcome $\phi \equiv E[Y_{d(a,A,W)}]$ corresponding to an intention-to-treat rule are straightforward to derive and are presented elsewhere [28]. In order to obtain a targeted MLE of $\phi$, we can use that by (6) the efficient influence curve of $\phi$ in a non-parametric model can be written as the sum of the efficient influence curve of a non-parametric estimator of $\phi_1 = E[YI(a \notin \mathcal{D})]$ and the efficient influence curve of a non-parametric estimator of $\phi_2 = E[Y_a I(a \in \mathcal{D})]$. These are given by

$$D^1(P) = I(a \notin \mathcal{D})Y - \phi_1 \tag{29}$$

and

$$D^2(P) = I(a \in \mathcal{D}) \left\{ \frac{I(A = a)}{g(A \mid W)} \Big[Y - Q(A, W)\Big] + Q(a, W) \right\} - \phi_2, \tag{30}$$

respectively, yielding

$$D(P) = I(a \notin \mathcal{D})Y + I(a \in \mathcal{D}) \left\{ \frac{I(A = a)}{g(A \mid W)} \Big[Y - Q(A, W)\Big] + Q(a, W) \right\} - \phi \tag{31}$$

as the efficient influence curve for $\phi$. The component of this influence curve corresponding to the score for $P(Y \mid A, W)$ is given by

$$\begin{aligned} D(P) &- E[D(P) \mid A, W] \\ &= I(a \notin \mathcal{D}) \Big[Y - Q(A, W)\Big] + I(a \in \mathcal{D}) \left\{ \frac{I(A = a)}{g(A \mid W)} \Big[Y - Q(A, W)\Big] \right\} \\ &= \left\{ I(a \notin \mathcal{D}) + I(a \in \mathcal{D}) \frac{I(A = a)}{g(A \mid W)} \right\} \Big[Y - Q(A, W)\Big]. \end{aligned} \tag{32}$$

The covariate $h(P_n^0)(A, W)$ needed for the univariate regression to update the initial fit for $Q$ is thus given by

$$h(P_n^0)(A, W) = I(a \notin \mathcal{D}) + I(a \in \mathcal{D}) \frac{I(A = a)}{g(P_n^0)(A \mid W)}. \tag{33}$$

The problems arising if the ETA assumption is violated are most clearly seen in the case of the IPTW estimator. By downweighting observations that were likely to have received their observed treatment and upweighting those that were instead unlikely to have received their observed treatment, this estimator essentially works by creating a new sample in which treatment assignment is independent of the baseline covariates. This approach breaks down if a subgroup of the target population never selects some of candidate treatment levels. If older, less healthy subjects, for example, are never observed to participate in high levels of vigorous physical activity, none of the subjects in the corresponding re-weighted sample will be older and less healthy, leading to an underestimate of the corresponding counterfactual mortality risk under high levels of vigorous physical activity.

In the same situation, the $G$-computation estimator has to rely entirely on model assumptions that cannot be tested from the observed data. Since older, less healthy subjects are never observed at higher levels of vigorous physical activity, their conditional mean outcome $E[Y \mid A, W]$ for these exercise levels is undefined. A corresponding estimate can never be obtained from the observed data unless one is willing to extrapolate from the conditional mean outcomes estimated for other values of $A$ and $W$. To illustrate this point, consider the simplified example in which $A$ is a binary indicator for a high level of vigorous physical activity and $W$ is an indicator for poor health. Then none of the subjects in our target population might fall in the group with $W = 1$ and $A = 1$ so that $E[Y \mid A = 1, W = 1]$ is undefined. In order to still obtain an estimate of this quantity, we would be forced to assume an additive model for $Q$ according to which $Q(A, W) = \beta_0 + \beta_1 A + \beta_2 W$. Since the non-parametric model $Q(A, W) = \beta_0 + \beta_1 A + \beta_2 W + \beta_3 A \times W$ is not identifiable, this assumption of no interaction between $A$ and $W$ cannot be tested from the observed data.

Like the $G$-computation estimator, the DR-IPTW estimator and the targeted MLE rely entirely on extrapolation through $Q$ if the ETA assumption is violated. To complicate matters, however, they also require that the estimate of $g$ is based on a model for the treatment mechanism that satisfies the ETA assumption, i.e. the model for $g$ must in fact be mis-specified [29]. In summary, all four estimators of causal effects are thus severely compromised if the ETA assumption does not hold, illustrating that the solution in such cases does not lie in turning to the $G$-computation or DR-IPTW estimators for which the resulting problems are not as immediately apparent as for the IPTW estimator, but in focusing on realistically defined causal effects that are guaranteed to be identified from the observed data.

## 4. Results

The treatment mechanism was estimated by a multinomial regression model that included main-effect terms for all indicator variables defined in Table 1. The regression $E[Y \mid A, W]$ was similarly estimated by a logistic regression model that included these same main-effect terms as well as indicator variables for the treatment categories 1 through 5. We evaluated the goodness-of-fit of this latter model using the Hosmer-Le Cessie test [7]. This test yielded a $p$-value of 0.10, providing little evidence against the assumption that this model adequately describes the data. To evaluate the fit of our treatment model, we followed the advise of Hosmer and Lemeshow [6] and treated this model as a set of independent binary logistic regression models of each treatment category against the remaining categories. Applying the Hosmer-Le Cessie test to each of these binary logistic regression models, we obtained $p$-values of 0.51, 0.54, 0.33, 0.27, 0.78, and 0.94, suggesting that the treatment model fits the data quite well.

Tables 2 and 3 summarize the fits we obtained for $g$ and $Q$, respectively. The treatment fit reveals that older, less healthy subjects do not have available the full set of treatment options: No subjects in the oldest age group (90-100 years) are observed at the treatment levels $A = 3$ and $A = 5$. Likewise, no subjects with poor self-rated health are observed at the treatment levels $A = 4$ and $A = 5$. In addition, subjects with severely impaired physical functioning are very unlikely to follow treatments $A = 4$ and $A = 5$. The fit we obtained for $Q$ indicates that these three groups of subjects are at considerably increased risks of mortality, suggesting that estimates of the counterfactual mortality risks for the higher three treatment categories will be biased low. Since the DR-IPTW estimator and the targeted MLE both require an estimate of the treatment mechanism that satisfies the ETA assumption, fitted treatment assignment probabilities below 0.05 were set to 0.05.

TABLE 2
*Treatment model fit. The entries in the first column give the factor by which the relative risk of falling in category A=1 rather than A=0 changes when the covariate under consideration is changed from 0 to 1. Entries in the remaining columns are interpreted accordingly.*

|            | A=1  | A=2  | A=3  | A=4  | A=5  |
|------------|------|------|------|------|------|
| AGE.1      | 1.16 | 1.57 | 1.37 | 1.32 | 1.44 |
| AGE.2      | 1.37 | 1.57 | 1.47 | 1.32 | 1.37 |
| AGE.4      | 0.74 | 0.94 | 0.83 | 0.83 | 1.02 |
| AGE.5      | 0.24 | 1.03 | 0.00 | 1.04 | 0.00 |
| HLT.EX     | 1.09 | 1.10 | 1.46 | 1.29 | 1.67 |
| HLT.FAIR   | 0.56 | 0.58 | 0.47 | 0.39 | 0.45 |
| HLT.POOR   | 0.50 | 0.43 | 0.33 | 0.00 | 0.00 |
| NRB.POOR   | 0.55 | 0.40 | 0.29 | 0.07 | 0.17 |
| NRB.FAIR   | 0.78 | 0.82 | 0.70 | 0.99 | 0.53 |
| SMOKE.CURR | 0.65 | 0.43 | 0.32 | 0.61 | 0.33 |
| SMOKE.EX   | 1.00 | 1.23 | 1.09 | 1.25 | 1.20 |
| CARD       | 0.90 | 1.29 | 1.18 | 0.89 | 1.46 |
| CHRONIC    | 1.19 | 1.14 | 1.13 | 1.11 | 0.93 |
| FEMALE     | 0.94 | 0.86 | 0.82 | 0.89 | 0.55 |
| DECLINE    | 0.67 | 0.39 | 0.52 | 0.37 | 0.33 |

Tables 4 and 5 summarize the realistic indvidualized treatment rule and the intention-to-treat rule. Both Tables show that only about 50% of all subjects are estimated to be capable of engaging in the highest level of vigorous physical activity. Likewise, only about 75% of all subjects are estimated to be capable of the second highest level. These observations further suggest that counterfactual outcomes for higher activity levels are not well defined for all subjects, or at least that a considerable porportion of subjects in the target population are rarely observed at these activity levels. In comparing Tables 4 and 5, we note that the intention-to-treat causal effects of high levels of vigorous physical activity are likely to be smaller than the corresponding realistic causal effects. Under the intention-to-treat rule $d(5, A, W)$, close to 25% of all subjects are assigned to the lowest treatment level $A = 0$ while the corresponding realistic individualized treatment rule $d(5, W)$ assigns no subjects to $A = 0$. In general, the realistic individualized treatment rule results in treatment assignments closer to the specified target level than those obtained from the intention-to-treat rule. In addition, the latter rule produces a few cases in which subjects are assigned to treatment levels that exceed the given target level. For the sake of estimating the causal effect of vigorous physical activity, these observations would seem to make the realistic individualized treatment rule a somewhat more appealing option than the intention-to-treat rule.

TABLE 3
*Fit for Q. Estimated odds ratios for mortality along with 95% confidence intervals and p-values.*

|  | OR | 95% CI | p-value |
|---|---|---|---|
| AGE.1 | 0.12 | (0.05, 0.31) | < 0.0001 |
| AGE.2 | 0.43 | (0.29, 0.64) | < 0.0001 |
| AGE.4 | 3.41 | (2.34, 4.96) | < 0.0001 |
| AGE.5 | 5.74 | (2.07, 15.91) | < 0.0001 |
| HLT.EX | 0.76 | (0.50, 1.16) | 0.2039 |
| HLT.FAIR | 2.01 | (1.39, 2.93) | < 0.0001 |
| HLT.POOR | 2.84 | (1.51, 5.34) | 0.0012 |
| NRB.POOR | 1.94 | (1.21, 3.13) | 0.0063 |
| NRB.FAIR | 0.89 | (0.61, 1.29) | 0.5279 |
| SMOKE.CURR | 3.73 | (2.22, 6.29) | < 0.0001 |
| SMOKE.EX | 1.38 | (0.99, 1.94) | 0.0584 |
| CARD | 1.60 | (1.13, 2.26) | 0.0080 |
| CHRONIC | 1.44 | (1.06, 1.95) | 0.0204 |
| FEMALE | 0.52 | (0.37, 0.72) | < 0.0001 |
| DECLINE | 1.46 | (1.05, 2.05) | 0.0266 |
| A=1 | 0.86 | (0.55, 1.34) | 0.5072 |
| A=2 | 0.81 | (0.51, 1.29) | 0.3849 |
| A=3 | 0.78 | (0.47, 1.29) | 0.3360 |
| A=4 | 0.45 | (0.18, 1.09) | 0.0770 |
| A=5 | 0.80 | (0.37, 1.76) | 0.5866 |

TABLE 4
*The realistic individualized treatment rule. A given row shows the treatment levels $\tilde{a} \equiv d(a, W)$ that subjects were actually assigned to when the target level was set at a.*

|  | $\tilde{a} = 0$ | $\tilde{a} = 1$ | $\tilde{a} = 2$ | $\tilde{a} = 3$ | $\tilde{a} = 4$ | $\tilde{a} = 5$ |
|---|---|---|---|---|---|---|
| $a = 0$ | 2051 | 0 | 0 | 0 | 0 | 0 |
| $a = 1$ | 11 | 2040 | 0 | 0 | 0 | 0 |
| $a = 2$ | 0 | 41 | 2010 | 0 | 0 | 0 |
| $a = 3$ | 0 | 41 | 97 | 1913 | 0 | 0 |
| $a = 4$ | 0 | 41 | 91 | 441 | 1478 | 0 |
| $a = 5$ | 0 | 41 | 91 | 381 | 454 | 1084 |

Table 5
*The intention-to-treat treatment rule. A given row shows the treatment levels $\tilde{a} \equiv d(a, A, W)$ that subjects were actually assigned to when the target level was set at $a$.*

|       | $\tilde{a}=0$ | $\tilde{a}=1$ | $\tilde{a}=2$ | $\tilde{a}=3$ | $\tilde{a}=4$ | $\tilde{a}=5$ |
|-------|------|------|------|------|------|------|
| $a=0$ | 2051 | 0    | 0    | 0    | 0    | 0    |
| $a=1$ | 11   | 2040 | 0    | 0    | 0    | 0    |
| $a=2$ | 35   | 3    | 2011 | 1    | 1    | 0    |
| $a=3$ | 108  | 16   | 7    | 1918 | 2    | 0    |
| $a=4$ | 338  | 88   | 66   | 56   | 1491 | 12   |
| $a=5$ | 492  | 161  | 134  | 110  | 45   | 1109 |

As argued above, the lack of non-parametric identifiability of causal parameters under a violation of the ETA assumption is most easily seen in the case of the IPTW estimator which is likely to suffer from considerable bias. Wang et al. [31] propose the following simulation-based approach for obtaining an estimate of this bias: Given estimates of $P(W)$, $g$, and $Q$, we can simulate realizations of the observed data structure. For this estimated data-generating distribution, the true parameter values for the parameters of interest can be computed through $G$-computation. At the same time, we can obtain a sampling distribution of IPTW estimates by applying the IPTW estimator to a large number of simulated realizations of the observed data structure. Since the assumption of no unmeasured confounders is trivially satisfied in this simulation study, any discrepancy between the mean of these estimates and the true parameter value must reflect a violation of the ETA assumption.

Table 6 summarizes the estimated bias of the IPTW estimator of the counterfactual mortality risk for each of the three different kinds of causal effects. The Table shows that the IPTW estimator dramatically underestimates the counterfactual mortality risk for static treatment interventions at the highest two activity levels, with considerable problems even for the third highest level of activity. We note that the estimated bias for the highest activity level should be treated with care since the corresponding static parameter is not even well defined. The remainder of the results is in agreement with our earlier arguments according to which a lack of older and less healthy subjects among the higher activity levels should lead to an underestimate of the corresponding mortality risks. In contrast, Table 6 shows only a negligible bias for estimating such risks on the basis of realistic individualized treatment rules and intention-to-treat rules. We stress that this diagnostic simulation should be interpreted to give not only an estimate of the bias seen in the IPTW estimator, but, more generally, a sense of the extent to which an ETA violation makes the causal parameters of interest non-parametrically non-identifiable or even ill defined. In the present case, for instance, we would therefore also want to treat any estimates of static causal effects offered by the $G$-computation, DR-IPTW, and targeted maximum-likelihood estimators as unreliable and potentially misleading.

Given the counterfactual mortality risk estimators described in section 3, estimators of the relative risk (relative to $A = 0$) are straightforward to obtain for the $G$-computation, IPTW, and DR-IPTW estimators by simply dividing the corresponding two mortality risk estimators. Since the targeted MLE is always aimed at a particular parameter of interest, this simple approach does not work for obtaining a targeted MLE of the relative risk of mortality. Section A in the appendix shows that this task is still fairly straightforward, however, given the work we have already done in section 3. Table 7 summarizes the relative risk estimates for the three different kinds of causal effects obtained by the four different estimators.

TABLE 6
*Estimated ETA bias for the IPTW estimator of the counterfactual mortality risk as a percentage of the true parameter value.*

|     | Static   | Realistic | ITT     |
| --- | -------- | --------- | ------- |
| A=0 | -0.23%   | -0.23%    | -0.23%  |
| A=1 | -2.63%   | 0.05%     | -0.03%  |
| A=2 | -4.94%   | 0.04%     | 0.13%   |
| A=3 | -14.45%  | 0.22%     | 0.20%   |
| A=4 | -48.75%  | 1.16%     | 1.05%   |
| A=5 | -50.54%  | -0.18%    | 0.11%   |

In the analysis based on static treatment interventions, the IPTW estimator appears to provide strong evidence for a protective effect of vigorous physical activity at the highest two levels, with an estimated 4-fold reduction in risk for the second-highest level. The realistic and intention-to-treat analysis, however, provide much weaker evidence for such a protective effect. As expected, the intention-to-treat causal effect estimates tend to be closer to the null value than the corresponding realistic estimates. Given the results of the simulation study summarized in Table 6, we are led to conclude that the IPTW estimates based on static treatment interventions dramatically overstate the beneficial impact of high levels of vigorous physical activity.

The remaining three estimators likewise tend to estimate stronger reductions in risk in the static analysis than in the realistic and intention-to-treat analyses, with both the DR-IPTW estimator and the targeted MLE indicating a significant protective effect for $A = 4$ in the static analysis that becomes non-significant in the realistic and intention-to-treat analyses. Interestingly, the $G$-computation estimator also yields a smaller estimated reduction in risk for $A = 4$ in the latter two analyses than in the former one, but tighter confidence intervals for the realistic and intention-to-treat analyses actually make the corresponding causal effect estimates significant while this is not the case in the static analysis. We speculate that the greater sampling variability observed in the static analysis is likely a result of the extrapolation that is required to estimate the expected mortality outcome for a large number of subjects that are never observed at the highest two treatment levels. For all four estimators, the static analysis suggest a markedly greater mortality risk for $A = 5$ than for $A = 4$, a finding that would be quite hard to interpret. The remaining two analyses, in contrast, provide much more compatible estimates for these two activity levels. These observations lend credence to the idea that the static effect estimates not only of the IPTW estimator, but also of the $G$-computation,

DR-IPTW, and targeted maximum-likelihood estimator ought to be treated as unreliable and potentially misleading. On the basis of the more trustworthy realistic and intention-to-treat analyses, the data suggest that high levels of vigorous physical activity may confer reductions in mortality risk on the order of 15-30%, although in most cases the evidence for such an effect does not quite reach the 0.05 level of significance.

TABLE 7
*Estimates of the relative risk of mortality (relative to $A = 0$) along with 95% confidence intervals based on the bootstrap.*

|  | G-comp | IPTW | DR-IPTW | tMLE |
|---|---|---|---|---|
| **Static** | | | | |
| A=1 | 0.90 (0.65, 1.20) | 0.97 (0.68, 1.29) | 0.96 (0.69, 1.28) | 0.96 (0.69, 1.28) |
| A=2 | 0.91 (0.64, 1.23) | 0.90 (0.60, 1.22) | 0.92 (0.63, 1.27) | 0.93 (0.63, 1.30) |
| A=3 | 0.88 (0.59, 1.21) | 0.77 (0.44, 1.07) | 0.84 (0.56, 1.14) | 0.87 (0.58, 1.18) |
| A=4 | 0.59 (0.22, 1.01) | 0.23 (0.06, 0.43) | 0.52 (0.20, 0.92) | 0.48 (0.15, 0.88) |
| A=5 | 0.86 (0.43, 1.35) | 0.55 (0.21, 0.90) | 0.97 (0.48, 1.50) | 1.05 (0.53, 1.60) |
| **Realistic** | | | | |
| A=1 | 0.91 (0.66, 1.19) | 1.00 (0.72, 1.32) | 0.95 (0.70, 1.28) | 0.95 (0.70, 1.28) |
| A=2 | 0.87 (0.63, 1.17) | 0.97 (0.67, 1.34) | 0.99 (0.66, 1.30) | 1.00 (0.66, 1.32) |
| A=3 | 0.85 (0.62, 1.13) | 0.81 (0.50, 1.22) | 0.91 (0.59, 1.22) | 0.91 (0.58, 1.23) |
| A=4 | 0.73 (0.53, 0.97) | 0.58 (0.34, 1.06) | 0.69 (0.40, 1.05) | 0.69 (0.41, 1.05) |
| A=5 | 0.81 (0.60, 1.06) | 0.66 (0.38, 1.19) | 0.78 (0.47, 1.17) | 0.78 (0.46, 1.20) |
| **ITT** | | | | |
| A=1 | 0.91 (0.66, 1.19) | 0.99 (0.72, 1.33) | 0.95 (0.70, 1.28) | 0.95 (0.69, 1.28) |
| A=2 | 0.88 (0.64, 1.17) | 0.98 (0.69, 1.31) | 0.98 (0.67, 1.29) | 0.98 (0.66, 1.30) |
| A=3 | 0.87 (0.64, 1.13) | 0.85 (0.59, 1.17) | 0.87 (0.61, 1.15) | 0.83 (0.60, 1.14) |
| A=4 | 0.78 (0.62, 0.97) | 0.85 (0.64, 1.08) | 0.84 (0.63, 1.04) | 0.85 (0.63, 1.10) |
| A=5 | 0.91 (0.75, 1.11) | 0.96 (0.73, 1.23) | 0.99 (0.73, 1.23) | 1.01 (0.73, 1.30) |

## 5. Discussion

The data analysis presented in this article illustrates the problems encountered in attempting to estimate the causal effect of a static treatment intervention if the ETA assumption is violated and some of the counterfactual outcomes of interest are not even well defined. While it is fairly well known that a violation of the ETA assumption can cause strong bias in the IPTW estimator, its effects on other estimators of static causal effects have received little attention in the literature. With the *G*-computation estimator, the DR-IPTW estimator, and the targeted MLE all relying on extrapolation from a correctly specified model for $Q$ and the latter two estimators in addition requiring a mis-specified model for the treatment mechanism that satisfies the ETA assumption, we argue that the results offered by these three estimators must also be treated with great caution. Since, strictly speaking, static causal effects cannot be identified from the observed data if the ETA assumption is violated, it should in fact make sense that the appropriate response to this problem does not lie in turning to approaches that aim to estimate such parameters by relying on untestable modelling assumptions, but rather in adapting the definition of the parameter of interest in a way that makes the parameter identifiable.

This becomes particularly obvious in cases in which static causal effects are not even well defined. In the context of studying the causal effect of vigorous physical activity on mortality in the elderly, for instance, it makes little sense to talk about the counterfactual outcome distribution we would observe if all subjects were assigned to high levels of activity since serious health problems would prevent a considerable proportion of subjects from complying with such an assignment. Causal effects defined on the basis of realistic individualized treatment rules and intention-to-treat rules address this problem by explicitly taking into account the set of treatment options that are realistically available to each subject. Such effects are therefore well defined and identifiable even if the full set of treatment options is not available to some subjects. The estimates of such effects reported here suggest that high levels of vigorous physical activity may confer reductions in mortality risk on the order of 15-30%, although in most cases the evidence for such an effect does not quite reach the 0.05 level of significance. Estimates of static causal effects, in contrast, suggest a statistically significant reduction in mortality risk on the order of 50-75%, a finding that given the estimated bias of the IPTW estimator, must be viewed as highly suspect.

A possible extension to the analysis we present here consists of data-adaptively selecting the value for $\alpha$ in definition (3) of the set of realistic treatment options, arbitrarily set by us as $\alpha = 0.05$. For very small values of $\alpha$, estimators of causal effects based on realistic individualized treatment rules and intention-to-treat rules may still be affected by a practical violation of the ETA assumption. As the value for $\alpha$ is increased, on the other hand, the corresponding causal effects become more and more different from the static causal effect that they are in some sense intended to approximate. A more sophisticated analysis might thus attempt to use the approach introduced by Wang et al. in order to find the smallest value of $\alpha$ for which the ETA bias of the IPTW estimator is estimated to be negligible. Future research will be required to investigate this approach further.

## 6. Acknowledgements

We would like to thank Dr. Ira Tager from the Division of Epidemiology at the UC Berkeley School of Public Health for kindly making available the dataset that was used in our data analysis. His work on the SPPARCS project was supported by a grant from the National Institute on Aging (RO1-AG09389).

## Appendix A: Targeted MLE of the causal relative risk

Let $\psi_a = E[Y_{d(a,W)}]$ and consider the parameter

$$\theta = \frac{E[Y_{d(a,W)}]}{E[Y_{d(0,W)}]} = \frac{\psi_a}{\psi_0}. \tag{34}$$

Since we have already derived the efficient influence curve of $\psi_a$ as

$$D^{\psi_a}(P) = \frac{I(A = d(a,W))}{g(A \mid W)}\Big[Y - Q(A,W)\Big] + Q(d(a,W),W) - \psi_a, \tag{35}$$

we can use the $\delta$-method to find the efficient influence curve for $\theta$. Specifically, we have that

$$\theta = f(\psi_a, \psi_0) = \frac{\psi_a}{\psi_0} \tag{36}$$

and

$$Df = (1/\psi_0, -\psi_a/\psi_0^2) \tag{37}$$

so that the efficient influence curve for $\theta$ is given by

$$\begin{aligned}
D(P) &= Df(D^{\psi_a}(P), D^{\psi_0}(P))^T \\
&= \frac{1}{\psi_0}\left\{\frac{I(A = d(a,W))}{g(A \mid W)}\Big[Y - Q(A,W)\Big] + Q(d(a,W),W) - \psi_a\right\} - \\
&\quad \frac{\psi_a}{\psi_0^2}\left\{\frac{I(A = d(0,W))}{g(A \mid W)}\Big[Y - Q(A,W)\Big] + Q(d(0,W),W) - \psi_0\right\} \\
&= \frac{1}{\psi_0}\Big[I(A = d(a,W)) - \theta I(A = d(0,W))\Big]\frac{Y - Q(A,W)}{g(A \mid W)} + \\
&\quad \frac{1}{\psi_0}\Big[Q(d(a,W),W) - \theta Q(d(0,W),W)\Big].
\end{aligned} \tag{38}$$

The component of this influence curve corresponding to the score for $P(Y \mid A, W)$ is given by

$$D(P) - E[D(P) \mid A, W] = \frac{1}{\psi_0}\Big[I(A = d(a,W)) - \theta I(A = d(0,W))\Big]\frac{Y - Q(A,W)}{g(A \mid W)}. \tag{39}$$

The covariate $h(P_n^0)(A, W)$ needed for the univariate regression to update the initial fit for $Q$ is thus given by

$$\begin{aligned} h(P_n^0)(A, W) &= \frac{I(A = d(a, W)) - \theta I(A = d(0, W))}{g(P_n^0)(A \mid W)\psi_0} \\ &= \frac{I(A = d(a, W)) - \psi_a/\psi_0 I(A = d(0, W))}{g(P_n^0)(A \mid W)\psi_0}. \end{aligned} \qquad (40)$$

To obtain a feasible $h(P_n^0)(A, W)$, we substitute

$$\psi_{a,n} = \frac{1}{n}\sum_{i=1}^n Q(P_n^0)(d(a, W_i), W_i) \qquad (41)$$

and

$$\psi_{0,n} = \frac{1}{n}\sum_{i=1}^n Q(P_n^0)(d(0, W_i), W_i) \qquad (42)$$

for $\psi_a$ and $\psi_0$, respectively. Let $\epsilon_n$ denote the MLE of $\epsilon$ in $Q(P_n^0)(\epsilon)$ and let

$$Q_n^1(A, W) = \frac{1}{1 + \exp(-m_n^0(W) - \epsilon_n h(P_n^0)(A, W))}. \qquad (43)$$

Iterate this process $k$ times until $\epsilon_n$ has become sufficiently small. Then the targeted MLE of $\theta$ is given by

$$\theta_n^{tMLE} = \frac{\sum_{i=1}^n Q_n^k(d(a, W_i), W_i)}{\sum_{i=1}^n Q_n^k(d(0, W_i), W_i)}. \qquad (44)$$

The covariate $h(P_n^0)(A, W)$ for the corresponding intention-to-treat relative risk parameter can similarly be derived as

$$\begin{aligned} h(P_n^0)(A, W) = {} & I(a \in \mathcal{D})\left[\frac{1}{\psi_0} - \frac{\psi_a}{\psi_0^2}\right] + \\ & I(a \notin \mathcal{D})\left[\frac{I(A = d(a, W)) - \psi_a/\psi_0 I(A = d(0, W))}{g_n^0(A \mid W)\psi_0}\right] \end{aligned} \qquad (45)$$